\documentclass[11pt]{article}

\newcommand{\s}{\rightarrow}

\usepackage{latexsym}
\usepackage{amssymb}
\usepackage{graphicx}
\newcommand{\x}{{\bf x}}
\newcommand{\y}{{\bf y}}

\newcommand{\IN}{{\cal N}}
\newcommand{\IY}{{\cal Y}}
\newcommand{\IF}{{\cal F}}

\newcommand{\R}{{\bf R}}

\newcommand{\Q}{{\bf Q}}

\newcommand{\PP}{{\bf P}}
\newcommand{\PR}{{\rm Prob}}

\begin{document}

\title{\hfill\mbox{{\small\it If you can look into the seeds of time,}}\\
\hfill\mbox{{\small\it And say which grain will grow, and which will not,}}\\
\hfill\mbox{{\small\it  Speak then to me.}}\\
\hfill\mbox{{\small W. Shakespeare, {\it
Macbeth}, I, 3.}}\\
\mbox{}\\
\mbox{}\\
{\bf Coins, Quantum Measurements, and  Turing's Barrier\thanks{A preliminary version
of this paper has appeared in \cite{boris}.}}}
\author{Cristian S. Calude\thanks{Department of Computer Science, The
University of Auckland, Private Bag 92019, Auckland, New Zealand.
E-mail: {\tt cristian@cs.auckland.ac.nz}.} \ and Boris
Pavlov\thanks{Department of Mathematics, University of Auckland,
Private Bag 92019, Auckland, New Zealand. E-mail: {\tt
pavlov@math.auckland.ac.nz.}}}
\maketitle \thispagestyle{empty}

\begin{abstract}
Is there any hope for quantum
computing to challenge the Turing barrier, i.e. to solve an undecidable
problem, to compute an uncomputable function? According to Feynman's '82
argument, the answer is {\it negative}.
This paper  re-opens the case: we will discuss solutions to a few simple
problems which
suggest that {\it quantum computing  is {\it theoretically}
capable of
computing uncomputable functions}. 

Turing proved that there is no ``halting (Turing) machine" 
capable of   distinguishing between halting and   non-halting programs (undecidability of the Halting Problem). Halting
programs can be recognized by simply running them; the main difficulty is to
detect non-halting programs. In this paper a mathematical quantum ``device" (with sensitivity $\varepsilon$) is constructed to solve the Halting Problem.
 The ``device" works on a randomly chosen test-vector  for   $T$ units of time.
If the ``device" produces a click, then the program  halts.
If it does not produce a click, then either the program does not halt
or the   test-vector has been chosen  from an  {\it undistinguishable 
set of vectors} ${\IF}_{\varepsilon, T}$. The last case is not dangerous as
our main result proves: {\it  the Wiener measure of} ${\IF}_{\varepsilon, T}$ {\it  constructively  tends to zero when} $T$ {\it tends to infinity}.  The  ``device",   working in time $T$, appropriately
computed,  will  determine
 with a pre-established precision whether an arbitrary program  halts or not. {\it Building the ``halting machine" is mathematically possible.}

To construct our ``device" we  use the quadratic form of an iterated map (encoding the whole data in an infinite superposition) acting on randomly chosen vectors viewed as special trajectories of two Markov processes working in two different scales of time. The evolution is described by an unbounded, exponentially  growing
semigroup;  finally  a single measurement produces the result.

\end{abstract}

\section{Introduction}
For over fifty years the Turing machine model of computation has defined
what it means to ``compute'' something; the foundations of the modern theory
of computing are based on it. Computers are reading text, recognizing speech,
and robots are
driving themselves across Mars. Yet this exponential race will not
produce solutions to many intractable and undecidable problems.
Is there any alternative? Indeed, quantum computing offers
one  such alternative (see \cite{casti97,criscasti,gruska,wc,cp}). To date, quantum
computing has been very successful in ``beating" Turing machines in
the race of solving intractable problems, with Shor and Grover algorithms
achieving the most impressive successes; the progress in quantum hardware is
also impressive. Is there any hope for
quantum computing to challenge the Turing barrier, i.e. to solve an undecidable
problem, to compute an uncomputable function? According to Feynman's
argument (see \cite{feynman82}, a paper reproduced also in
\cite{feynman99}, regarding the possibility of simulating
a quantum system on a (probabilistic) Turing machine\footnote{Working with
probabilistic Turing machines instead of Turing machines makes no difference in
terms of computational capability: see \cite{delemosh}.}) the answer is {\it negative}.

\medskip

This paper re-opens the case:\footnote{See \cite{cds,cp,etesi,kieu2} for related
ideas and results.} We will discuss solutions to a few 
simple problems
which
suggest that {\it quantum computing is theoretically capable of computing
uncomputable functions}.
The main features of our  quantum ``device" are:
{\it
a special type of continuity, 
the  choice of  test-vectors from a  special
class of trajectories of  two Markov processes 
  working in  two different scales of time
and realized as elements of an infinitely-dimensional Hilbert
space
 (infinite superposition),
  the ability to work with ``truly random" test-vectors
 in an  evolution  described by  an exponentially  growing
semigroup and the possibility to obtain the result  from
a  single  measurement.}

\medskip

In deciding
the halting/non-halting status of  a non-halting  machine,  our  ``device"  is capable to `announce' (with a
positive probability) the non-halting  status in a finite amount of time, well before  the `real' machine reaches it (in an infinite amount of time).
Hence,  the
challenge was to design a procedure that detects and measures this  tiny, but
non-zero probability. 
\medskip

In what follows
a {\it quantum}  solution is a solution designed to work on a
quantum computer. The discussion is {\it mathematical} and no engineering claims
will be made; in particular, when speaking about various quantum devices which will be
constructed, we will use quotes to emphasize the mathematical nature of our constructs.

\section{The Merchant's Problem}
One possible way to state the famous Merchant's Problem
\if01
\footnote{The problem has circulated during
the WW2. One of the authors (B.P.)  learnt it  from retired Russian colonel
A. N. Ignatov in 1950.}
\fi
is as follows:

\begin{quote}
\em A merchant learns than one of his five stacks of $\Gamma =1$ gram coins
contains only false coins,  $\gamma = 0.001$ grams heavier than normal
ones. Can he find the odd stack by a single ``weighting"?
\end{quote}

\noindent The well-known  solution  of  this  problem  is the
following:
   We  take  one  coin from  the
first  stack, two  coins
  from  the  second stack, \ldots,  five  coins  from  the  last  stack.

\medskip

Then
  by measuring  the  weight  of  the  combination  of coins described above
   we obtain the number $ Q = 15 + \gamma \times n $
  grams ($1 \leq n \leq 5$), which  tells  us that  the  $n$-th  stack
contains
  false  coins.

\medskip

The above  solution  is,  in spirit,  ``quantum".
It consists of the following steps: a)
{\it preparation}, in which a single object encoding the answer
of the problem is created in a special format, b) {\em
measurement}, in which a measurement is performed on the object, c)
{\it classical calculation}, in which
the result produced is processed and the desired final result is obtained.

\bigskip\begin{center}
\includegraphics[width=2in]{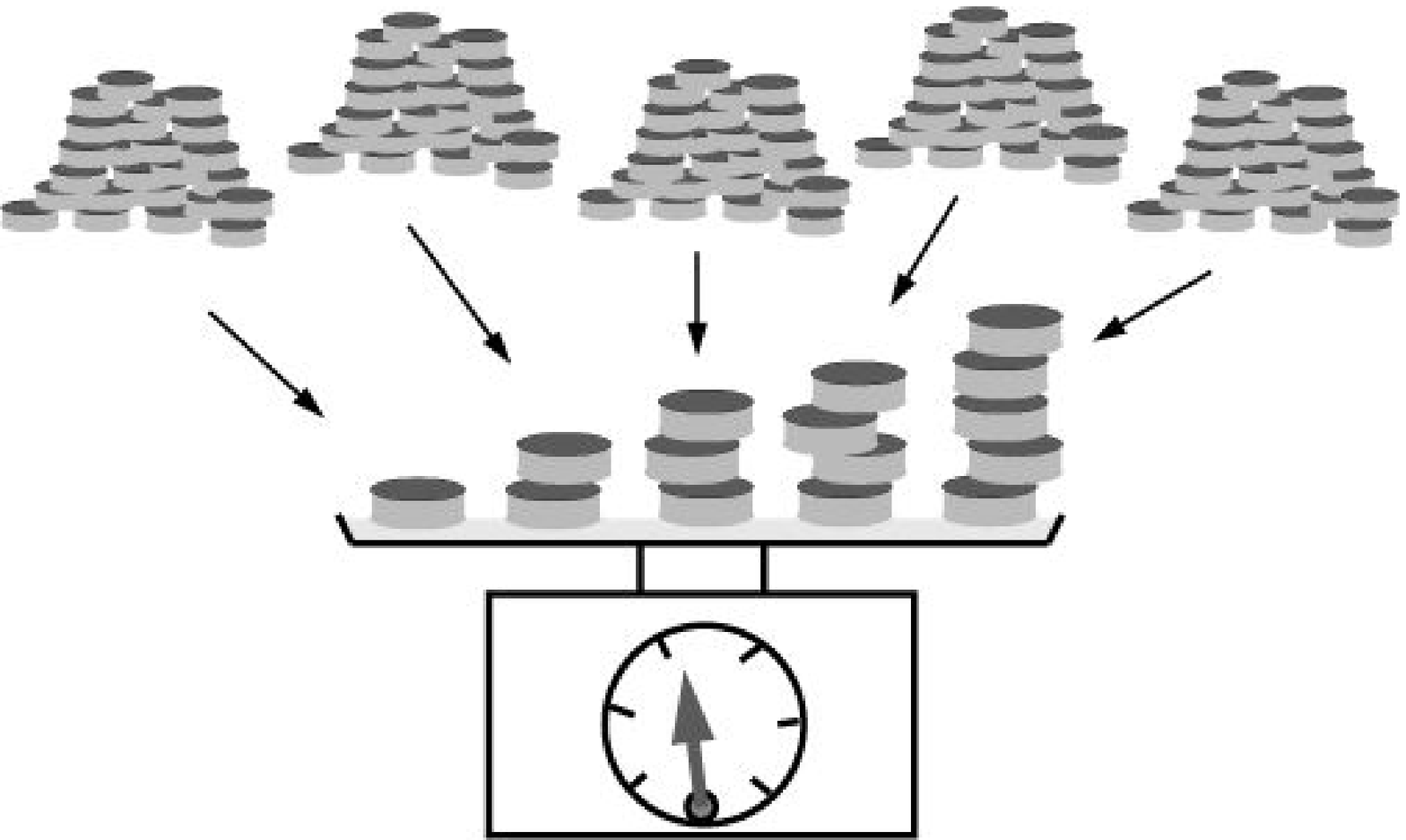}

\bigskip

Figure 1. Coin selection
\end{center}
\bigskip

In our case, the selection of coins from various stacks as
presented in Figure 1 is the object a) prepared for measurement
b); finally, the calculation $n = (Q-15) \times 1000$ gives the
number of the stack containing false coins.

\section{The Merchant's Problem: Two  Finite Variants}

Consider now the case when we have  five stacks
of coins, but  a few (maybe none) may contain false coins. This means,  all five stacks contain true coins, or only one stack
contains false coins, or  two stacks contain false coins, etc.
Can we, again with only one single ``weighting", find all
stacks containing false coins? A possible solution is to
choose  1, 2, 4, 8, 16 coins from each
stack, and use the uniqueness of base two representation.

\medskip

The difference between the above solutions
is only in the specific way we chose the sample, i.e.
in {\it coding}.
Further on, note that the above solutions  work {\em only} if
we have {\em enough coins in each stack}. For example, if each of the five
stacks
contains only four coins, then neither of the above solutions
works. In such a case is it still possible to have a solution operating with
just one measurement?

\medskip
In the simplest case  we have $N$ stacks of coins and we know that {\it
at most one stack may contain false coins}. We are  allowed
to take just one coin from each stack and we want to see whether
all coins are true or there is a stack of false coins. Can we solve 
this problem
with just one ``weighting"?

\medskip

Assume that a true coin has $\Gamma = 1$ grams and a false coin has
$\Gamma + \gamma$ grams ($0 < \gamma < 1$).
Consider as quantum space the space $H_N =\R^{N}$, a real Hilbert space of
dimension $N$.
The elements of
$\R^N$ are vectors $\x = (x_1, x_2, \ldots ,x_N)$. The scalar product of $\x$ and $
\y$
is defined by $\langle \x, \y \rangle = \sum_{i=1}^N x_i y_i$.
The norm of the vector $\x$ is defined by $\parallel \x \parallel  =
\sqrt{\langle
\x,
\x\rangle}$. Let
$0<n<N$, and consider $\Omega^{n} \subset \R^n$. A  set $X \subset \R^N$ is
called {\it cylindrical} if $X = \Omega^n \times \R^{N-n}$. Let
us denote by $\mu^k$ the Lebesgue measure in $\R^k$. If  $\Omega^n 
\subset \R^n$
is measurable, then the cylinder $X = \Omega^n \times \R^{N-n}$
is measurable and
  $\mu^N (X) = \mu^n (\Omega^n)$.
  For  more on Hilbert spaces see
\cite{ag,halmos}; for specific relations with quantum physics see \cite{cohen}.

\medskip
Next we consider the standard basis $(e_i)_{i = 1, N}$ and the projections
$\PP_i:
\R^N \s \R^N$, $\PP_i (\x) = (0,0,\ldots ,x_i, 0,\ldots ,0)$. Denote by $q_i$
the weight of a coin in the $i$-th stack; if the $i$-th stack contains true
coins,
then $q_i = \Gamma = 1$, otherwise, $q_i = \Gamma + \gamma= 1 + \gamma$.

\medskip

Consider the operator  ${\bf Q} = \sum_{i=1}^{N} q_i \PP_i$.\footnote{As suggested by  \cite{radu},
different
operators can be considered, e.g.
${\bf Q}(x)= \sum_i^N 2^{(q_i- \Gamma)} \, \PP_i$.}  For every
vector
$\x\in\R^N$,

\[
{\bf Q}(\x)  =  (q_1 \PP_1, \ldots , q_N \PP_N)(\x) = (q_1x_1, \ldots , q_N x_N).\]

  The $t$-th ($t>1$)
iteration of the operator ${\bf Q}$ can be used to distinguish the 
case in which
all
coins are true from the case in which one stack contains false coins: we
construct
the quadratic form $\langle {\bf Q}^{t}(\x), \x
\rangle$ and consider its dynamics.\footnote{To speed-up the computation one can accelerate the iterations of  ${\bf Q}$, for example by considering 
the quadratic form  $\langle {\bf Q}^{2^t}(\x), \x
\rangle$ instead of $\langle {\bf Q}^{t}(\x), \x
\rangle$.} In case all coins are
true
$\langle {\bf Q}^{t}(\x), \x \rangle = \, \parallel \x \parallel^2$, for all $\x\in\R^N$; if there
are false coins in
some stack, for some $\x\in\R^N$,
$\langle {\bf Q}^{t}(\x), \x \rangle > \, \parallel \x \parallel^2$, and the value increases with every
new
iteration.

\medskip

Now we can introduce  a
   ``weighted  Lebesgue  measure" with  proper non-negative
continuous density  $\rho$.
For example, this can be achieved with the  density equal to the
Gaussian distribution
$$\rho (\x) =  \frac{1}{\pi^{N/2}} \,\, e^{-\sum_{s = 1}^N
|x_s|^2},$$ a function which will be used in what follows.

\medskip

We can interpret the measure generated by the density as the probability distribution
corresponding to the standard  $Normal \, (N; 0,\frac{1}{2}{\bf I})$.  Hence  the  probability  of  the  event
$
\{\x
\mid x_1 \in \Omega\}$ is the  integral $\PR (\Omega) = \int_{\Omega \times \R^{N-1}} \rho dm.$
Then,
because of the continuity of the  density, we deduce that
the  probability  of  any ``low-dimensional event" is  equal  to
zero.  In particular, the  event  $\{\x \mid x_s = 0\}$  has 
probability zero, that is, with probability one all components of a randomly
chosen normalized vector $\x$ are non-zero.

\medskip

We are now ready to consider our problem. We will assume that time is discrete,
$t
= 1,2, \ldots$. The procedure will be {\it probabilistic\/:}  it will indicate a
method
to decide, with a probability as close  to
one
as we
want,  whether there exist any false coins.

\medskip
Fix a computable real $\eta \in (0,1)$ as probability threshold.  Assume that
both   $\eta$ and
$\gamma$ are computable reals.
Choose  a ``test" vector $\x \in \R^N$. Assume that we have a quantum
``device"\footnote{The construction
of such a ``device" is a difficult problem in nanoelectronics; see, for example,
\cite{euro,boris1,boris2}.} which
measures the quadratic form and clicks at time
$T$ on
$\x$ when

\begin{equation}
\label{click}
\langle {\bf Q}^{T}(\x), \x
\rangle > (1 + \varepsilon) \parallel \x \parallel ^2.
\end{equation}

\noindent In this case we say that the quantum ``device" has sensitivity
$\varepsilon$.
In what follows we will assume that $\varepsilon > 0$ is a positive computable
real.

\medskip

Two cases may appear. If for some $T>0$, $\langle {\bf Q}^{T}(\x), \x \rangle >
(1
+ \varepsilon) \parallel \x \parallel ^2,$ then the ``device" has clicked and we
know
for {\em sure} that there exist false coins in the system. However, it is
possible
that at some time $T>0$ the ``device"  hasn't (yet?) clicked because $\langle {\bf
Q}^{T}(\x), \x \rangle \le (1 + \varepsilon) \parallel \x \parallel ^2.$ This
may
happen because either all coins are true, i.e.,
$\langle {\bf Q}^{t}(\x), \x \rangle = \, \parallel \x \parallel^2$, for all $t>0$, or because at time
$T$ the growth of $\langle {\bf Q}^{T}(\x), \x \rangle $ hasn't yet reached the
threshold $(1 + \varepsilon) \parallel \x \parallel ^2$. In the first case the
``device" will {\em never} click, so at each stage $t$ the test-vector $\x$
produces
``true" information; we can call $\x$ a  ``true"  vector. In the second case, the
test-vector $\x$ is ``lying" at time
$T$ as we {\it do}  have false coins in the system, but they were not detected
at time
$T$; we say that $\x$ produces ``false" information at time $T$.

\medskip

Hence, the ``true"
vector  has non-zero coordinates corresponding to  stacks of false coins (if
any);
a vector  ``lying" at time
$T$  may have zero or small coordinates corresponding to  stacks of 
false coins.
For instance, the null vector produces ``false" information at any time. If the
system has false coins and they are located in the $j$-th stack, then each test
vector $\x$ whose  $j$-th coordinate is 0 produces ``false" information at any
time.
If the system has false coins and they are located in the $j$-th stack, $x_j
\not= 0$, but
$$ \parallel \x \parallel ^2 + ((1+\gamma)^T -1) |x_j|^2 \le (1 + \varepsilon)
\parallel \x
\parallel ^2,$$ then
$\x$ produces ``false" information  at time $T$. If $|x_j| \not= 0$, then
$\x$ produces ``false" information  only a finite period of  time, 
that is, only
for
$$T \le \log_{1+\gamma} \left( 1 + \frac{ \varepsilon \parallel \x \parallel
^2}{|x_j|^2}\right);$$ after this time the quantum ``device"  starts clicking.

\medskip

The major problem is to distinguish between the presence/absence of false
coins in the system. We will show how to compute
the time $T$ such that  when presented
a randomly chosen test-vector\footnote{A different approach would be to consider the (constant) test
vector $ \x= (\frac{1}{\sqrt{N}}, \frac{1}{\sqrt{N}}, \ldots ,\frac{1}{\sqrt{N}})$            playing the role of an equal ``superposition" of all stacks.} $\x \in \R^N \setminus \{{\bf 0}\}$
to a quantum ``device" with sensitivity $\varepsilon$ that fails to click in time
$T$,
then  the system doesn't contain false coins with  probability larger than
$1-\eta$.

\medskip

Assume first that the system contains    false  coins in some stack $j$. Then

\begin{equation}
\label{inf}
\lim_{t\s \infty}  \frac{\langle \Q^{t}(\x),\x \rangle}{\parallel \x \parallel
^2}
= \infty,
\end{equation}
for all $\x \in \R^N$ such that $|x_i|\not= 0$, for all $1\le i \le N$. Indeed,
in view of the hypothesis, there exists $j \in \{1, 2, \ldots ,N\}$ such that
the
weight of any coin in the $j$-th stack, $q_j$,  is $\Gamma + \gamma = 1 +
\gamma$.
So, for every $t\ge 1$,

\[\langle \Q^{t}(\x),\x \rangle
= \sum_{i=1}^N q_i^t \parallel \x \parallel ^2\\
= \, \parallel \x \parallel ^2  +  ((1 + \gamma)^t -1) |x_j|^2.\]

If $|x_j| \not= 0$, for all $j \in \{1, 2, \ldots ,N\}$, then

\[\lim_{t\s \infty}  \frac{\langle \Q^{t}(\x),\x \rangle}{\parallel 
\x \parallel
^2}
=  \lim_{t\s \infty}  1 + \frac{((1 + \gamma)^t -1) |x_j|^2}{\parallel \x
\parallel ^2} = \infty.
\]

If the system contains only true coins, then for every $\x \in \R^N \setminus
\{{\bf 0}\},$
$$\lim_{t\s \infty}  \frac{\langle \Q^{t}(\x),\x \rangle}{\parallel \x
\parallel ^2} = 1.$$

Consider now the {\it indistinguishable set at time}  $t$

\[ {\IF}_{\varepsilon, t} = \{\x \in \R^N \mid \langle \Q^{t}(\x),\x \rangle
\le (1 + \varepsilon) \parallel \x
\parallel ^2 \}.\]

If the system contains only true coins, then ${\IF}_{\varepsilon, t} = \R^N$, for
all
$\varepsilon >0, t \ge 1$. If there is one stack (say, the $j$-th one)
containing false
coins, then  ${\IF}_{\varepsilon, t}$ is a cone ${\IF}_{\varepsilon, t, j} $ 
centered at
the
``false" plane $x_j = 0$:

\[ ((1+\gamma)^t - 1) \, |x_j|^2 \, \le \,\, \parallel x \parallel^2.\]

Next we  compute $\PR({\IF}_{\varepsilon,
t})$ in case the {\it system contains false coins}.
Each set ${\IF}_{\varepsilon, t} = {\IF}_{\varepsilon, t, j}$ can be 
decomposed into two
disjoint sets as
follows (here $M>0$ is  a large enough real which will be determined later):

\[ {\IF}_{\varepsilon, t, j} = \{\x \in {\IF}_{\varepsilon, t, j} \mid M \ge \parallel
\x
\parallel  \} \cup
\{\x \in {\IF}_{\varepsilon, t, j} \mid   M <  \parallel \x
\parallel  \}.\]

In view of the inclusion

\[ \{\x \in {\IF}_{\varepsilon, t, j}\mid M\ge \, \parallel \x
\parallel \}  \subset \{\x \in \R^N \mid ((1+\gamma)^t - 1) \, |x_j|^2
\le \varepsilon M^2\},  \]

\noindent we deduce that

\begin{eqnarray}
\label{est1}
\PR (\{\x \in {\IF}_{\varepsilon, t, j}  \mid M \ge \, \parallel \x \parallel  \}) &
\le &
\frac{1}{\sqrt{\pi}}
\int_{-\frac{M\sqrt{\varepsilon}}{\sqrt{(1+\gamma)^t - 1}}}
^{\frac{M\sqrt{\varepsilon}}{\sqrt{(1+\gamma)^t -
1}}} \,\, e^{-y^2} dy \nonumber \\
& \le &
\frac{2 M\sqrt{\varepsilon}}{\sqrt{\pi} \sqrt{(1+\gamma)^t -
1}},
\end{eqnarray}

To estimate $\PR(\{\x \in {\IF}_{\varepsilon, t} \mid M < \, \parallel \x
\parallel  \})$ we note that the set

\[ C_M = \bigcup_{i=1}^N \{\x \in
\R^N
\mid |x_i|> \frac{M}{\sqrt{N}}\}, \]

contains  the set  $\{\x \in {\IF}_{\varepsilon, t} \mid  M < \, \parallel \x \parallel  \},$
hence from the estimation

\[\PR(C_M) \le  \frac{2N}{\sqrt{\pi}} 
\int_{\frac{M}{\sqrt{N}}}^{\infty} \, e^{-
y^2}
dy,
\]

we deduce (using the inequality $\int_{a}^{\infty}\, e^{-y^2} dy \le
\frac{1}{2a}
\,\, e^{-a^2}$ for $a > 0$) that

\begin{equation}
\label{est2}
\PR (\{\x \in \R^N \mid M < \, \parallel\x \parallel, |x_j| \le
\frac{M}{\sqrt{N}}\}) \le \frac{N\sqrt{N}}{M\sqrt{\pi}}\,\,
e^{-\frac{M^2}{N}}.
\end{equation}

    From (\ref{est1}) and  (\ref{est2}) we obtain the inequality:

\begin{equation}
\label{est3}
\PR({\IF}_{\varepsilon, t}) = \PR({\IF}_{\varepsilon, t, j})\le \frac{2  M
\sqrt{\varepsilon}}{\sqrt{\pi}
\sqrt{(1+\gamma)^t - 1}} + \frac{N \sqrt{N}}{M \sqrt{\pi}}\,\, e^{-
\frac{M^2}{N}}.
\end{equation}

Selecting

\[M = N^{3/4} \cdot \left(\frac{1+\gamma)^t -
1}{\varepsilon}\right)^{1/4},\]

\noindent in (\ref{est3}) we get\footnote{Lemma~4  in  \cite{gelfand}, p. 325-326,
can be used to obtain a similar, but less tight estimation; cf. \cite{Lodkin}.}  

\begin{equation}
\label{est4}
\PR({\IF}_{\varepsilon, t})  \le
\frac{3 N^{3/4} \varepsilon^{1/4}}{\sqrt{\pi}((1+\gamma)^t-1)^{1/4}}
\end{equation}

\noindent hence,
\[
\lim_{t \s \infty} \PR({\IF}_{\varepsilon, t})   =
0.\]

The above limit is {\em constructive},  that is, from  (\ref{est4}) and  every
computable
$\eta
\in (0,1)$ we can construct the computable bound
\begin{equation}
\label{eta}
T_{\eta} = \log_{1+\gamma} \left(   \frac{3^4 N^{3} \varepsilon}{
\eta^4 \pi^2}    + 1\right)
\end{equation}
\noindent such that {\it
assuming that the system contains false coins, if} $t \ge T_{\eta},$
{\it then  we get}
$$\PR({\IF}_{\varepsilon,t}) \le \eta.$$

Recall that we have a finite system of $N$ stacks in which at most one stack
contains
false coins. So, if we assume that there are $N+1$ equiprobable possibilities,
then either all coins are true or
only the first  stack
contains false coins, or only the second  stack
contains false coins, or only the $N$th  stack
contains false coins.\footnote{Of course, other distributions can be considered.}  Let
us now denote by
$\IN$ the event ``the system contains no false coins" and by $\IY$ the event
``the system contains
false coins". By $ P (\IN)$  ($P (\IY)$) we denote the {\it a priori}
probability that the
system contains no false coins (the system contains
  false coins). In the simplest case   $P (\IY) = \frac{N}{N+1}, P  (\IN)
= 1 - P(\IY) = \frac{1}{N+1}$.  We can use Bayes' formula to obtain the {\it a
posteriori
probability that the system contains only true coins when at time $t$ the
quantum ``device" didn't click}:

\[
P_{\mbox{non-click}} (\IN) = \frac{P(\IN)}{P (\IN) +
(1- P(\IN)) \PR({\IF}_{\varepsilon,t})} \ge 1 - N \cdot \PR({\IF}_{\varepsilon,t}).
\]

When $t \rightarrow \infty$,  $\PR(\Omega_{\varepsilon,t})$ goes to $0$, so
$P_{\mbox{non-click}}
(\IN)$ goes to  $1$.  More precisely, if $t \ge T_{\eta},$ as in (\ref{eta}),
then

$$P_{\mbox{non-click}}
(\IN) \ge 1 - \eta N.$$

In conclusion,
\begin{quote}
\it for every computable $\eta \in (0,1)$ we can construct a computable time
$T_{\eta}$ such that
picking up at random a test-vector $\x \in \R^N \setminus \{{\bf 0}\}$  and
using a
quantum ``device" with sensitivity $\varepsilon$ up to time
$T_{\eta}$ either

\begin{description}
\item  $\diamond \,$
    we get a click at some time $ t \le T_{\eta}$, so the system
contains false coins, or
\item $\diamond \,$  we don't get a click in time $T_{\eta}$, so  with
probability greater than
$1-\eta N$ all coins are true.
\end{description}
\end{quote}

\section{The Merchant's Problem: The Infinite Variant}
Let  us  assume that
  we  have  now a  countable  number  of stacks, all   of  them, except at most
one,
  containing   true  coins  only. Can we determine whether there is
a stack containing false coins?
  It is not difficult to recognize
that the infinite variant of the Merchant's Problem is equivalent to 
the Halting
Problem: Decide whether
an arbitrary program (Turing machine, probabilistic Turing machine, Java
program, etc.) eventually
halts. This problem is undecidable, i.e., no Turing machine can solve it.\footnote{Arguably, the most famous  undecidable problem. See, for
example,
\cite{cris}.}  

\medskip

One of the most important quests of science is to determine those (natural) processes whose final state may be determined directly, without a need to exhaustively carry out each step of their evolutions. Usually, this is done by a  ``model"  that ``simulates" the process. The essence of the undecidability of the Halting Problem is the following: If our models are only Turing machines,  then the outcome of the computation performed by a  Turing machine can, in general, be determined {\it only } by explicitly carrying out each step of it. No short-cut is possible. Can we do it better if we enlarge the class of models?  We shall prove that this is indeed the case.

\subsection{A Tentative Solution}
The first idea would be to follow the solution discussed in  Section 
3,  but to  select the random  test
  vector $\x = (x_0,\,x_1,\, x_3,\dots)$ from  the  Hilbert  space
  $H = l_2$ of  quadratically summable   
sequences of  probabilistically independent
         variables $x_i$,   equipped
with the
  Gaussian distribution on all cylindrical
sets with finite-dimensional sections parallel to coordinate planes.  
\medskip

 We define $$\langle \Q^{T}(\x),\x \rangle
= \sum_{i=1}^{\infty} q_i^T |x_i|^2.$$

The analogue of the
{\it indistinguishable set}  in
$l_2$
\if02
  corresponding
  to  the  simplest  ``device"  used  for  measurement of  the  exponentially
  growth of  the false  component of  the  randomly  chosen
    vector   $Q^T \x,\,\, T \to \infty $ was  defined as  a  set
  of  all vectors $x$ for  which
\fi
is

\[{\IF}_{\varepsilon, T} = \{\x \in \l_2 \mid \langle \Q^T (\x), 
\x\rangle \leq (1
+ \varepsilon)
\langle
\x,
\x\rangle \} \]
  \begin{equation}
  \label{device}
\phantom{xxxxxx} = \{\x \in \l_2 \ \langle \Q^T (\x), \x\rangle \leq
\langle \x, \x\rangle + \langle \varepsilon {\bf I} (\x), \x\rangle\}.
  \end{equation}

\noindent  so, the    measuring  ``device" is
   the  operator  $\varepsilon I$.
If  for a given test-vector $\x$  we  have
$\langle Q^T (\x), \x\rangle \geq \, (1 + \varepsilon) \parallel \x \parallel^2 $  ($
\parallel \cdot \parallel$ is the $l_2$--norm), then  the ``device"  clicks,
which  means that  there  is  a  false  coin  in  some  stack   $j$
(represented  by  a  non-zero  component $x_j$  of  the  test-vector  $\x$).
  If  the  ``device"  does  not  click, then  the  result  of  the  experiment
  is  not  conclusive:  either  we  do  not  have  false  coins  in 
the  system,
  or, we have, but  the  test  vector  ``lies"  since  it  belongs to
the set  ${\IF}_{\varepsilon, T}$  of  indistinguishable  elements.

\medskip

Assume that the system contains false coins in some stack $j$. For large $T$ such that $(1+\gamma)^T > 1 + \varepsilon$, the   coordinate description  of 
the  set 
${\IF}_{\varepsilon, T}$  can be  given  in  the form of  a  cone  centered  at  the
``false plane" $x_{j} = 0$ in $H$:

\[
{\IF}_{\varepsilon, T} = \left\{\x\mid |x_{j}|^2 \leq
\frac{\varepsilon}{(1+\gamma)^T -1} \, \parallel \x \parallel ^2
\right\}.
\]

Consider now the intersection of  the  indistinguishable  set  ${\IF}_{\varepsilon, T}$
with the  finite-dimensional subspace  $H_{2n} = \{{\bf x} \mid x_i = 0, i> 2n
\}$,  ${\cal F}_{\varepsilon, T, 2n}= {\cal F}_{\varepsilon, T}\cap
H_{2n}$.
It is clear that ${\IF}_{\varepsilon, T, 2n}  \subset {\IF}_{\varepsilon, T,
2n+1}.$
Let $\varepsilon \cdot ((1 + \gamma)^T -1)^{-1}$ be denoted by $\alpha^2$.
Assume for a moment that the Gaussian distribution may be extended
 by  Lebesgue  procedure to a  probability  measure Prob. Then, we can
 calculate the measure of the finite-dimensional
 section ${\cal F}_{\varepsilon,T,N}$ of  the indistinguishable
 set ${\IF}_{\varepsilon, T}$ (if $N=2n$):

   \[\mbox{Prob}({\IF}_{\varepsilon, T,N}) =
\frac{
\int_0^{\alpha \sqrt{n}}\frac{dv}{(1+ v^2/n)^{n}}
}{
\int_0^{\infty} \frac{dv}{(1+ v^2/n)^{n}}
}.
  \]

In view  of the Lebesgue dominant convergence theorem
($\int_{0}^{A}
\frac{dv}{(1+v{^2}/^n)}
\rightarrow \int_{0}^{A} e^{-v^2}dv$)  the  limit of
 $\mbox{Prob}({\IF}_{\varepsilon, T,N})$
  can be estimated as follows: when
$n \to \infty$, $\mbox{Prob} ({\IF}_{\varepsilon, T, 2n}) \to \frac{\int_0^{\alpha
\sqrt{n}}{e^{-v^2}dv}}
      {\int_0^{\infty}         {e^{-v^2}dv}
}$ uniformly in $\alpha,
\, 0 <\alpha <\infty$, and

  \begin{equation}
\label{apprestim}
\frac{\int_0^{\alpha \sqrt{n}}{e^{-v^2}dv}}
      {\int_0^{\infty}         {e^{-v^2}dv}
}  =  \frac{2}{\sqrt{\pi}}\int_0^{\alpha \sqrt{n}} {e^{-v^2}dv}.
  \end{equation}

  If the duration of the experiment is fixed ($T$ is constant), but $n$
tends  to infinity,
then the   measure Prob$({\IF}_{\varepsilon, T, 2n})$ of  the
finite-dimensional indistinguishable set ${\IF}_{\varepsilon, T, 2n}$ tends to
1. Hence, in view of the assumption on Prob,   monotonicity   and  the inclusion 
${\IF}_{\varepsilon, T, 2n} \subset {\IF}_{\varepsilon, T} $  we
conclude  that Prob$({\IF}_{\varepsilon, T}) = 1, $  for all $T$, hence
  $\lim_{T\to\infty}$ Prob$({\IF}_{\varepsilon, T}) = 1 $.

\medskip

 On  the  other
hand, ${\IF}_{\varepsilon, T'}\subset {\IF}_{\varepsilon, T}$,  if $T' > T$ and
$\bigcap_{T>0}{\IF}_{\varepsilon, T} = \lim_{T\to \infty} {\IF}_{\varepsilon,
T} = \{\x \mid x_j = 0\}$ is a cylindrical set  with   measure 0.
This implies that  our  assumption about the possibility  to  construct
 the  Lebesgue  extension  of  the Gaussian  distribution   was  wrong.
This  is  the  mathematical reason  why  our  ``device" will work only `locally', on the observed finite part of the system, not `globally', on the whole infinite system.

\medskip

  Assume  that  we  are  dealing  with a  class  of  systems  where  the  {\it a
priori} probability  of  absence  of  false  coins  is    $P (\IN)$.
We  select at random  one  of  these  systems  and  perform    experiments   using 
our  ``device". Then, due  to  Bayes'  formula,  the {\it a posteriori} probability  of 
absence  of  false  coins  in  the  system  {\it subject  to the  assumption that  the 
``device"  did  not  click  in time }  $T$ is 

\[
P_{\mbox{non-click}} (\IN)  = \frac{P (\IN)}{P (\IN) + (1- P (\IN))
\mbox{Prob}({\IF}_{\varepsilon, T})},
\]

\noindent so if $\mbox{Prob}({\IF}_{\varepsilon, T}) = 1$, then  
$$P_{\mbox{non-click}} (\IN) = P(\IN),$$
  hence  the  ``non-click" result   is  not  conclusive.
Still,  formula (\ref{apprestim})  suggests  a procedure
for  estimating  the  {\it a posteriori} probability  of
presence  of  false  coins  in  the  {\it   observed} finite  part  of
the system.

\medskip
Assume  that we have observed  the first    $2n$  elements of    the
system.   Further, suppose that the  duration of  the  experiment  $T$ and
the above number $n$ satisfy  
  the following  condition:  

\begin{equation}
\label{nt}
\alpha \sqrt{n} = \sqrt{\frac{\varepsilon n}{(1 + \gamma)^T - 1}} \longrightarrow  0,
\end{equation}

\noindent when $n \to \infty$. Let $ \Gamma (n) = \alpha \sqrt{n}$.  Then,  according  to  
(\ref{apprestim}) 
   we  have:

\[
\lim_{n \to \infty} \mbox{Prob} ({\IF}_{\varepsilon, T, 2n}) = \lim_{n \to \infty}
\frac{1}{\sqrt{\pi}} \int_0^{\Gamma (n)}e^{- x^2} dx =  0.
\]

Hence,  using again  
Bayes's  formula, if   $T \to  \infty$  and  $T, n$ satisfy (\ref{nt}), then

\[
P_{\mbox{non-click}} (\IN) = \frac{P (\IN)}{P (\IN) + (1- P (\IN))
\frac{1}{\sqrt{\pi}} \int_0^{\Gamma (n)}e^{- x^2} dx} \longrightarrow  1,
\]

\noindent when $n\to\infty$.

Because  of  the  revealed  ``discontinuity"  of the
 Gaussian  distribution  in  $l_2$,\footnote{Lack of   countable
            additivity of its extension.}   
 the probability of  the high-dimensional sections of the indistinguishable set  (\ref{device}) 
is  not uniformly  small in $n$,
  for  large $T$.  This  is  in
   agreement  with  the   view that
``only  a
finite  number of subjects may  be  observed in  finite  time".\footnote{According to
Theorem~2   
in \cite{gelfand}, p. 345,    the 
Lebesgue  extension  of  the
Gaussian  measure   in a countably  Hilbert  space  exists   if  and  only  if    the
distribution function   is    equal to
$e^{-\langle Ax,x \rangle}$,
where  $A$ is a  Hilbert-Schmidt operator. If the condition is not satisfied,  then    the 
Lebesgue  extension
of  the  Gaussian  measure still   exists, but  in  a  larger  Hilbert  space, 
in which   the  initial
Hilbert  space  has  measure  zero.}
In fact, the problem is related to  the mathematical notion of finiteness, which appears to be  ``inadequate to the task of telling us which
physical processes are
finite and which are infinite" (see \cite{del}).

\subsection{A Brownian Solution}

The failure  of the tentative approach was   caused by  the
  structure of  the  stochastic  space  of  
 test-vectors. A  more elaborated  approach, developed in this section,  will permit the  estimation of
   the probability of absence of  false coins in  the
whole {\it infinite} sequence by observing the behaviour  of the quadratic form
of the iterated map

$$\langle \Q^t (\x), \x\rangle
= \sum_{i=1}^{\infty} q_i^t  \, |x_i|^2$$

\noindent   on randomly chosen test-vectors $\x$
viewed  {\it  as   special   trajectories}  of a Markov  process.\footnote{As
in the finite case, various other choices of operators  can be considered in order to speed-up the computation.}

\medskip

To this aim we drop the assumption of probabilistic independence and consider a  ``device"  detecting  the  false  coins
which  is based on   {\it continuous} probability  measures  induced by     Markov 
processes, see  \cite{Belop90}. We  construct  two  Markov
processes   
working in {\it two different
discrete time  scales}. To capture the idea of ``continuity" referred to in Sections~1 and 4.1   the construction makes use of  the Green function of the
  Cauchy problem for the heat equation

\begin{equation}
\label{heat} \frac{\partial G}{\partial t}= \frac{1}{4}
\frac{\partial^2 G}{\partial x^2},\,\,\,\,\, G(x,y,0)= \delta
(x-y),
\end{equation}

\noindent which  may  be  interpreted  (see, for example,   \cite{Freidlin})
as  a probability--density of the space--distribution of a Brownian
particle on  the  real axis which  begins  diffusion from the
initial position $y$ at the initial moment  $t=0$:

  \begin{equation}
  \label{Green}
  G (x,t\big|y,0)= \frac{1}{\sqrt{\pi t}} e^{-\frac{|x-y|^2}{t}}.
  \end{equation}

  The  Green  function is  a positive  analytic  function of each
  variable in  the  half-plane $0<t<\infty,\,\,\, -\infty < x<
  \infty$. It    provides   information on  the
  distribution of the  Brownian  particle  on  the  whole
  infinite  axis {\it for any positive time} $t>0$, which  corresponds
  to   diffusion with    {\it infinite  speed}. 

\medskip

We are going to use three  spaces. The first is  
the    stochastic space of  all  trajectories  $\x$ of
Brownian particles   equipped  with the Wiener
measure  $W$ (see \cite{stroock}).  The measure $W$ is 
defined  on the  algebra  of  all finite-dimensional cylindrical sets
$C_{_{_{\Delta_1,\Delta_2,\dots ,\Delta_{N}}}}^{t_1,t_2,\dots
,t_{N}}$ of  trajectories with fixed  initial  point
$x_0= 0$ and ``gates" $\Delta_l,\, l = 1, \ldots, N$ (which are
open intervals on the real line):

\[
C_{_{_{\Delta_1,\Delta_2,\dots ,\Delta_{N}}}}^{^{^{t_1,t_2,\dots
,t_{N}}}} = \left\{ \x \mid x_{t_l}\in \Delta_l, l=1,2,\dots ,N
\right\},
\]

\noindent via multiple convolutions  of the Green  functions
$G(x_{l+1},t_{l+1}\big| x_l,t_l) $  corresponding
to  the   steps $\delta_{l+1} = t_{l+1}- t_l$:

  \[
W^N (C_{_{_{\Delta_1,\Delta_2,\dots
,\Delta_{N}}}}^{^{^{t_1,t_2,\dots ,t_{N}}}}) =
\phantom{xxxxxxxxxxxxxxxxxxxxxxxxxxxxxxxxxxx}
\]  \\[-10ex]

\begin{equation}
\displaystyle
  \label{cylinder} \frac
  { \int\dots\int_{\Delta_{N},\Delta_{N-1},\dots
,\Delta_{1}} \frac{dx_1 dx_2 \dots dx_{N}}{\pi^{\frac{N}{2}
\sqrt{ \delta_{N} \delta_{N-1}\dots \delta_1 }}}
\, e^{^{-\frac{|x_N - x_{N-1} |^2}{\delta_N}}}\dots e^{^{-\frac{|x_1
- x_{0} |^2}{\delta_1}}} }{ \int\dots\int_{\R_{N}, \R_{N-1},\dots
,\R_{1}} \frac{dx_1 dx_2 \dots dx_{N}}{\pi^{\frac{N}{2} \sqrt{
\delta_{N} \delta_{N-1}\dots \delta_1 }}} \, e^{^{-\frac{|x_N -
x_{N-1} |^2}{\delta_N}}}\dots e^{^{-\frac{|x_1 - x_{0}
|^2}{\delta_1}}} },
\end{equation}

\noindent where $\R_{N} = \R_{N-1} = \dots
= \R_{1} = \R$. Using the convolution formula, 
the
denominator of   (\ref{cylinder})  can  be  reduced to the  Green  function 
 $ G(x_N,t_N \mid 0, 0)$,
for  any  $\tau \in (s,t)$:

\[  G(x,t \mid y,s) = \int_{-\infty}^{\infty}  G(x,t \mid \xi,\tau)
  G(\xi,\tau \mid y,s) d\xi.
\]

Our   ``device"
(with sensitivity $\varepsilon$) will distinguish the values of
the iterated quadratic forms by observing  the difference between the
non-perturbed  and perturbed sequences $t_l,\,
  \tilde{t}_l$.
Instead  of  the  Hilbert  space $l_2$ we  will work with  its
  intersections with the  discrete Sobolev class  $l_2^1$  of
summable sequences  with  the square norm 

$$ \mid  \x 
\mid^2_1 \,\,\,
= \, \sum_{_{_{m = 1}}}^{\infty}
\,|x_{_m} -
x_{_{m-1}}|^2,  $$

\noindent and  the  discrete Sobolev class  $\tilde{l}_2^1$  of
weighted-summable sequences  with  the square norm 

$$  \parallel  \x 
\parallel ^2_1 \,\,\,
= \, \sum_{_{_{m = 1}}}^{\infty}
\frac{1-\tilde{\delta}_m}{\tilde{\delta}_m}\,\,\,|x_{_m} -
x_{_{m-1}}|^2.  $$

  We consider   two discrete stochastic processes corresponding to the {\it equidistant
sequence} of moments of  time  $ t_l = l ,\,\,l=0,1,\dots,\,\, \delta_s = 1$
and to the {\it perturbed sequence} of moments of  time $\tilde{t}_l = \sum_{m=0}^l
\tilde{\delta}_m,\,\, \tilde{\delta}_m < 1$.  We  assume  that $\tilde{t}_l$ are computable
and  for large values of $m$,  $\sum_m (1-\tilde{\delta}_m) < \infty$,  that is
$$\tilde{t}_N = N -  \sum_{m =1}^{N} (1-\tilde{\delta}_m) = 
N \left( 1 - \frac{\sum_{m =1}^{N} (1-\tilde{\delta}_m)}{N}\right) \approx N,$$ 
\noindent
for  large  $N$.
By natural extension from cylindrical sets we can define   the Wiener
measures  $\tilde{W}$ and $W$   on these spaces. In what follows we are going to use the
following relation between  $\tilde{W}$ and $W$  (see \cite{stroock}): for every
$W$--measurable set
$\Omega$,

\begin{equation}
\label{relW}
\tilde{W}(\Omega) = \frac{1}{\prod_{l=1}^{\infty} \sqrt{\delta_l}}
\int_{\Omega} e^{- \sum_{m=1}^{\infty} \frac{1-\tilde{\delta}_m}{\tilde{\delta}_m} \mid x_m -
x_{m-1}
\mid^2 } dW.
\end{equation}

Further we consider  the class  of
 {\it quasi-loops}, that  is the class of  all  trajectories  of  the  perturbed
process   which begins  from
$ (x_0, t_0) = (0,0 )$  and there exists a constant $C$ such that
$\max_{0< s <t} |x_s|^2 < Ct$. We note that
{\it 
\begin{itemize}
\item every $\x \in l_2^1$ is a quasi-loop (with $C = \,  \mid \x
\mid_1^2$),
\item due to the reflection principle {\rm (see \cite{stroock}, p. 221)}, the class of
all quasi-loops has Wiener measure one.
\end{itemize}
}

\medskip

We  assume that  our  ``device" cannot identify the
false coin at time $T$ in case the
test  vector $\x$ belongs  to  the {\it indistinguishable set}

\begin{eqnarray*}
\IF_{\varepsilon, T} &  = & \{ \x \in l_2 \cap l_2^1 \mid  \langle \Q^t (\x), \x \rangle
< \,
\parallel \x \parallel^2 \\ [2ex]
&&+ \,\, \varepsilon\,\, \left(  \sum_{_{_{m =
1}}}^{\infty}
\frac{1-\tilde{\delta}_m}{\tilde{\delta}_m}|x_{_m} -
x_{_{m-1}}|^2\right)\}\\ [2ex]
&  = & \{ \x \in l_2 \cap l_2^1 \mid  \langle \Q^t (\x), \x \rangle
< \,
\parallel \x \parallel^2  + \,\, \varepsilon 
\parallel\x \parallel^2_1 \}.
\end{eqnarray*}

If we assume that there exist false coins in the system, say at stack $j$, then 
\[\IF_{\varepsilon, T} = \{ \x \in  l_2^1 \mid  ((1+\gamma)^T-1) \mid x_j \mid^2 <
\, \varepsilon 
\parallel\x \parallel^2_1, \, \mbox{for some   } j\}.
\]

Next we
  will show that the  {\it  Wiener measure of the indistinguishable set
$\tilde{W}({\IF}_{\varepsilon,T})$  converges constructively to  zero when  $T\to \infty$.}
More precisely, we are going to prove that

\begin{equation}
\label{fin}
\tilde{W}({\IF}_{\varepsilon,T})\le
\left(\frac{\varepsilon} {
((1+\gamma)^T -1-\varepsilon) \cdot \prod_{m=1}^{\infty}\, \tilde{\delta}_m
}  \right)^{\frac{1}{2}}.
\end{equation}

\if01

\frac{\sqrt{\varepsilon}}{\sqrt{(1+\gamma)^T -1-\varepsilon} \, \sqrt{\prod_{m=1}^{\infty}\tilde{\delta}_m}}.

To this aim we will show
that  the measure
$\tilde{\mu_0}(\IF_{\varepsilon, T })$, associated with the
discrete Markov process labeled by the perturbed  time intervals
$\tilde{\delta}_l$, can be estimated  as
\begin{equation}
  \label{fin}
\tilde{\mu_0}\left(\IF_{\varepsilon, T} \right) <
\prod_{l=1}^{\infty} \frac
{1}{\sqrt{\tilde{\delta}_l}}\frac{\sqrt{\varepsilon
}}{\sqrt{\frac{\varepsilon}{j} +  ((1 + \gamma)^T -1) }} <
\prod_{l=1}^{\infty} \frac
{1}{\sqrt{\tilde{\delta}_l}}\frac{\sqrt{\varepsilon}}{\sqrt{(1 +
\gamma)^T -1}}.
  \end{equation}

The proof of the above estimate (\ref{fin})
is based on the reduction of the calculation of both the numerator and the
denominator of the expression for the measure of the indistinguishable set
corresponding to the similar integrals on the non-perturbed  process.
\par
  Really, the  numerator  of  the  fraction (\ref{tmeasure}) may  be
transformed on the finite-dimensional  sections $\Omega_M$
of  the  set $\Omega$ the  following  way:
\[
  \int\dots\int_{\Omega_M}{\pi^{\frac{M}{2}}
\sqrt{\tilde{\delta}_{M} \tilde{\delta}_{M-1}\dots \tilde{\delta}_1 }}
e^{^{-\frac{|x_M - x_{M-1} |^2}{\tilde{\delta}_M}}}\dots e^{^{-\frac{|x_1
- x_{0} |^2}{\tilde{\delta}_1}}}=
\]
\[
\frac{1}{\sqrt{\tilde{\delta}_{M} \tilde{\delta}_{M-1}\dots \tilde{\delta}_1 }}
  \int\dots\int_{\Omega_M} \frac{dx_1 dx_2 \dots dx_{M-1}}{\pi^{\frac{M}{2}}}
e^{^{-\sum_{m=1}^{m=M}\frac{|x_m - x_{m-1} |^2}{1}}}
  e^{^{-\sum_{m=1}^{m=M} (1 - \tilde{\delta}_m)\frac{|x_m - xm-1_
|^2}{\tilde{\delta}_m}}},
\]
and  the  denominator  may  be  also  reduced  to  the  form  containing
the  characteristics of  the  non-perturbed process:
\[
G(x_M,\tilde{t}_M \big| 0,0) = \frac{1}{\sqrt{\pi \tilde{t}_M}}
e^{^{ - \frac{|x_M|^2}{\tilde{t}_M}}} =
\]
\[
\frac{1}{\sqrt{\pi M}} \frac{\sqrt{M}}{\sqrt{\tilde{t}_M}}
e^{^{-\frac{|x_M|^2}{M}}} e^{^{ \frac{|x_M|^2 (M -
\tilde{t}_M)}{M\tilde{t}_M}}}.
\]
Then  using the  fact  that $\frac{M}{\tilde{t}_M} \to 1$ if $M \to \infty$
  we  may  see  that  for  $M \to \infty$
\[
  G(x_M,\tilde{t}_M \big| 0,0) =
\]
\[\frac{1}{\sqrt{\pi M}}
e^{-\frac{\mid x_M \mid ^2}{M}}
\left\{
\frac{1}{\sqrt{1-\frac{\sum_{m=1}^{M}(1-\tilde{\delta}_m)}{M}}}
e^{-\frac{\mid x_M
\mid ^2}{M^2}
\left(\frac{\sum_{m=1}^{M}(1-\tilde{\delta}_m)}
{1-\sum_{m=1}^{M}\frac{1-\tilde{\delta}_m}{M}}\right)}\right\}
\]
\[\approx \frac{1}{\sqrt{\pi M}}e^{-\frac{\mid x_M \mid^2}{M}} \]

if $\bf{x}$ is a  quasi-loop  beginning  from the  origin,
$x(0)=0$.

Then the limit of the ratio (\ref{tmeasure})
  may be calculated as an integral on the corresponding
Wiener measure $\mu_0$ associated with the non-perturbed discrete
  equidistant sequence of  moments of  time  $t_l, \,\, t_m - t_{m-1} =1$.
\[
\tilde{\mu}_0 (\Omega) =
\frac{1}{\sqrt{\prod_{l=1}^{\infty} \tilde{\delta}_l}}
\int_{\Omega}e^{-\sum_{m=1}^{\infty}\frac{1-\tilde{\delta}_m}
{\tilde{\delta}_m}\mid
x_m - x_{m-1} \mid^2} d{\mu}_0.
\]
The functional $e^{-\sum_{m=1}^{\infty}\frac{1-\tilde{\delta}_m}
{\tilde{\delta}_m} \mid  x_m - x_{m-1} \mid^2}:
=e^{-F}$ is bounded and hence the measure $\tilde{\mu}_0$
associated with the perturbed process is absolutely continuous with respect to
that associated with the non-perturbed process.

Now we estimate the measure of the
  indistinguishable set $\IF_{\varepsilon, T }$ taking into account that on
this set

\[
e^{-F} \le e^{-\frac{(1+\gamma)^T-1}{\varepsilon}\mid x_j \mid^2} \]

\fi

We now have:

\[\tilde{W} ({\IF}_{\varepsilon, T })
\phantom{xxxxxxxxxxxxxxxxxxxxxxxxxxxxxxxxxxxxxxxxxxxxxxx}
\]\\[-7ex]
\begin{eqnarray*}
& \le &
\frac{1}{\sqrt{\prod_{l=1}^{\infty} \tilde{\delta}_l}} \, \sup_k \, 
\int_{\mbox{quasi-loops}} \,\,
e^{-\frac{(1+\gamma)^T-1}{\varepsilon}\mid x_k \mid^2} dW\\
&=& \frac{1}{\sqrt{\prod_{l=1}^{\infty} \tilde{\delta}_l}}\, \sup_k \,  \lim_{C\to
\infty}\lim_{N \to \infty}
\\
&&
\frac{\int_{|x_N| < C\sqrt{N}}\int_{-\infty}^{\infty}
G(x_N,N\mid  \xi,k) \,
e^{-\frac{(1+\gamma)^T-1}{\varepsilon}\mid \xi \mid^2}
G(\xi,k\mid  0,0) \, dx_N d\xi}
{\int_{|x_N| < C\sqrt{N}} G(x_M,\, M \mid 0,0) \, dx_N} \\
& = & 
\frac{1}{\sqrt{\prod_{l=1}^{\infty} \tilde{\delta}_l}}
  \,  \sup_k \,  \lim_{C\to \infty}\lim_{N \to \infty} \frac{\sqrt{\pi N }}
{\pi\sqrt{k(N-k)}}\\
&&\frac{\int_{-C\sqrt{N}}^{C\sqrt{N}}
\int_{-\infty}^{\infty} e^{^{ - \frac{|\xi|^2}{k} - \frac{(1+\gamma)^T-1}{\varepsilon}\mid \xi
\mid^2 - \frac{\mid x_N - \xi \mid^2}{N-k} }} d\xi dx_N}
{\int_{-C\sqrt{N}}^{C\sqrt{N}} e^{^{-\frac{\mid x_{N} \mid^2}{N}  }} dx_N}
\end{eqnarray*}

The inner integral in the numerator may be explicitly calculated as:

$$ \int_{-\infty}^{\infty} e^{-\left( \frac{1}{k}+
\frac{(1+\gamma)^T-1}{\varepsilon} +
\frac{1}{N-k}\right) \xi^2} \, e^{2\frac{\xi x_N}{N-k}}\, e^{-\frac{1}{\mid
N-k\mid}\mid x_N \mid^2}\, d\xi
  \]
$$ = \frac{e^{-\frac{1}{\mid N-k \mid}
\mid x_N \mid^2}\sqrt{\pi}e^{\frac{\mid x_N \mid^2}{\mid N-k
\mid ^2}\frac{1}{\left(\frac{1}{k}+
\frac{(1+\gamma)^T-1}{\varepsilon}+
\frac{1}{N-k}\right)}}}{\sqrt{\frac{1}{k}+
\frac{(1+\delta)^T-1}{\varepsilon}+\frac{1}{N-k}}}.
\]

The  integrated
exponential  in  the  numerator  becomes:

\begin{eqnarray*}
 e^{- \frac{|x_N|^2}{N-k}\left( 1 - \frac{1}{(N-k)(\frac{1}{k}+
 \frac{(1+\gamma)^T - 1}{\varepsilon} + \frac{1}{N-k})} \right)}
& = &
e^{- \frac{|x_N|^2}{N-k}\left( 1 - \frac{1}{\frac{N}{k}+
 \frac{(1+\gamma)^T - 1}{\varepsilon} } \right)} \\
& = & e^{- \frac{|x_N|^2}{N-k}} \, e^{ \frac{|x_N|^2}{N-k}
 \left(\frac{1}{\frac{N}{k}+
 \frac{(1+\gamma)^T - 1}{\varepsilon} } \right)}\\
& = & e^{- \frac{|x_N|^2}{N-k}} \, e^{ \frac{|x_N|^2}{N-k}
 \left(\frac{\varepsilon}{\frac{N}{k}\varepsilon +
 (1+\gamma)^T - 1 } \right)}\\
& < & e^{- \frac{|x_N|^2}{N-k}} \, e^{ \frac{|x_N|^2}{N-k}
 \left(\frac{\varepsilon}{
 (1+\gamma)^T - 1} \right)}\\
& = & e^{-
\frac{|x_N|^2}{N-k}\left( 1 -
 \frac{\varepsilon}{(1+\gamma)^T - 1} \right)}.
\end{eqnarray*}

Finally, in view of the relation

\[\lim_{C\to\infty} \int_{-C\sqrt{N}}^{C\sqrt{N}} \frac{e^{- \frac{|x_N|^2}{N}}  \, dx_N}{\sqrt{N}}=
\lim_{C\to\infty} \int_{-\sqrt{C}}^{\sqrt{C}} e^{- v^2} dv = \sqrt{\pi},  \]

\noindent we obtain the  estimation  (\ref{fin}) of  the  measure  of  the
indistinguishable  set

\begin{eqnarray*}  W({\IF}_{\varepsilon, T}) 
&\leq &
\frac{\sqrt{\varepsilon}} {\sqrt{(1+\gamma)^T -1}}
\frac{1}{
\sqrt{ \prod_{m=1}^{\infty}\, \tilde{\delta}_m }
\, \sqrt{1 - 
    \frac{\varepsilon}{(1+\gamma)^T -1)}}
}\\
& = &
\left(\frac{\varepsilon} {
((1+\gamma)^T -1-\varepsilon) \cdot \prod_{m=1}^{\infty}\, \tilde{\delta}_m
}  \right)^{\frac{1}{2}}.\end{eqnarray*}

For example, if we put
\[  T_{\eta} = \log_{1+\gamma} \left(\frac{\varepsilon}{\eta^2 \, \prod_{m=1}^{\infty}\,
\tilde{\delta}_m}+ 1+ \varepsilon\right),\]
\noindent then $\tilde{W}({\IF}_{\varepsilon,T})\le \eta$ provided $t > T_{\eta} $.  For
example, if
$\tilde{\delta}_m = e^{-2^{-m}}$, for all $m\ge 1$, then  $T_{\eta}  = \log_{1+\gamma}
(\varepsilon e \eta^{-2}+1+\varepsilon)$.

\medskip

To conclude our analysis, we use  Bayes' formula in (\ref{fin}) to estimate
  the probability of absence of  false  coins  in  the
   system  when the ``device" does not click in time $T$ 
on randomly  chosen  test-vectors  selected  from the  class  of  quasi-loops. Using the same notation as in the end of Section~3, we have

  \[
P_{\mbox{non-click}} (\IN)>  1 - \frac{1- P(\IN)}{P(\IN)} \cdot
\frac{\sqrt{\varepsilon}}{\sqrt{(1+\gamma)^T-1-\varepsilon} \, \sqrt{\prod_{m=1}^{\infty}\, \tilde{\delta}_m}}.\]

\bigskip

\section{Is the Brownian Solution ``Quantum"?}

It is now the time to ask ourselves the question: Is the method ``quantum" or not?
After all, as one referee  and \cite{radu} have pointed out, ``continuous evolution in time and
space \ldots \/ is a common property of physical systems, classical as well as quantum".

\medskip

Not surprisingly, our approach goes, in a sense, beyond the ``classical" model of quantum computing in which a quantum Turing machine is the prototype.\footnote{See, for example, \cite{gruska,cp}. A similar remark can be made for the approaches discussed in \cite{etesi,kieu2}.} 
A quantum Turing machine is a straightforward generalization of a Turing machine, in which the main ingredients are (a) (entangled) qubits that can be in  various superpositions  (b) a universal  set of one-qubit and two-qubit unitary gates. It is designed to construct large, but {\it finite} unitary operations that can speed up the classical computation, say, by using   quantum  {\it finite} parallelism. By ``default" these models cannot cope  with the task
of solving an undecidable problem. The new ingredients built in our ``device"
include the use of an infinite superposition (in an infinite-dimensional  Hilbert space) which creates an
 ``infinite
type of quantum parallelism" and  the ability to work with ``truly random" vectors
 in an  evolution  described by  an exponentially  growing
semigroup.

\medskip

At this stage the ``device" is more mathematical than
physical.
To simplify the formalism  we have   used a
  real  Hilbert  space  (which  is  not  typical  for  
  quantum   problems)  because (a) it   supports the  superposition
  principle and (b) has the typical  features  of
  quantum computing. The method is essentially  quantum because
we code the whole (infinite) data in an infinite superposition (the  Hilbert  space),
we  assume that we have the ability to generate ``truly random"  vectors in the
Hilbert space and finally we 
apply  one single  measurement (via  the  quadratic  form) to obtain the result.
The method was  inspired by and is closer ``in spirit" to Benioff and Feynman early works \cite{Benioff80,feynman82}.

\medskip

An  essential question concerns  the  type  of
evolution. The  evolution we  use  is a  semigroup,
more  precisely, an  unbounded, exponentially  growing
semigroup.  The ability
of extracting the required (finite) information from an infinite data in a finite amount of time is given in part by the ``huge"  growth of this semigroup.\footnote{Compare  with the following paragraph from \cite{feynman65}: ``It bothers me that, according to the laws as we understand them today, it takes a computing machine an infinite number of logical operations to figure out what goes on in no matter how tiny a region of space, and no matter how tiny a region of time. {\it How can all that be going on in that tine space\/?} Why should it take an infinite amount of logic to figure out what a tiny piece of space-time is going to do?}
Clearly, this  is  not the 
typical evolution  for  ``quantum"  systems;  it is not difficult, but tedious (see \cite{ak,boris3}),    to  transform this evolution into  an equivalent  unitary one.\footnote{This will be the object of   a  separate  paper.}

\section{Final Comments}
We have discussed a few simple problems and their solutions in the quest of
finding a quantum approach for an undecidable problem. To this aim we have chosen
the infinite variant of the Merchant's problem which is equivalent to the
Halting Problem, the most well-known undecidable problem.

\medskip

 Halting
programs can be recognized by simply running them; the main difficulty is to
detect non-halting programs.
In deciding
the halting/non-halting status of  a non-halting  machine,  our  ``device"  is capable to `announce' (with a
positive probability) the non-halting  status in a finite amount of time, well before  the `real'  machine reaches it (in an infinite amount of time). The device  detects and measures this  tiny, but
non-zero probability. The method (described in Section~4.2) uses a  quadratic form of an iterated map (encoding the whole data in an infinite superposition) acting on randomly chosen vectors viewed as special trajectories of two Markov processes working in  two different scales of time.\footnote{Various natural ideas
fail to produce
exactly the desired result; one of them was discussed in Section~4.1.}

\medskip

The methods for trespassing Turing's barrier discussed by both 
Etesi and
N\'{e}meti 
\cite{etesi}
and Kieu \cite{kieu2}, although drastically different, have been, in some sense, prefigured by the accelerated Turing machines first imagined by Hermann Weyl
(see, for example, the discussion in Svozil \cite{karl}). The main task of their authors is not to describe their methods, but to argue/prove that they do not contradict any
{\it known} physical law. If a method would be shown to not be ``theoretically implementable", then the result would still be interesting as that would  show  a new type of computational limit, {\it physical}, not logical. 

\medskip

In our case, the main result is {\it mathematical}. We have proved that
{\it  the Wiener measure of the indistinguishable set} ${\IF}_{\varepsilon, T}$ {\it  constructively  tends to zero when} $T$ {\it tends to infinity}.  The  ``device",   working in time $T$, appropriately computed,  will  determine
 with a pre-established precision whether an arbitrary program  halts or not. {\it Building the ``halting machine" is mathematically possible.}
\medskip

 The   discrete-time  Brownian  motion--used
in the estimation   of   the   probability  of  the 
indistinguishable  set  in   the  last
  section--can be represented as   a
``sum"  of   independent  random   variables   with  Gaussian 
distributions.  It can be implemented as a ``sum" of spins
of a cascade of electrons formed by the shock-induced emission on a special
geometrical structure of semiconductor elements with special random properties
(cf. \cite{Yafyasov}).

\medskip

Many problems are still open and much more remains to be done.  Is  the  method 
used in this
paper ``natural"? Is it feasible?\footnote{See also \cite{cds}.}
 Is it better  or can we get more ``insight" about the nature of the Halting Problem if we use unitary operators?

\medskip

The results discussed in this paper, as well as  \cite{cds,cp,etesi,kieu2},  go beyond the pure
mathematical aspects;  they  might impose the re-examination of  the mind--machine issue
(see \cite{jack}).

\section*{Acknowledgements}
  Ya.~Belopolskaya \cite{yana} has suggested  the book \cite{stroock}  and the use
of the reflection principle, and M.~ Dumitrescu \cite{monica} and R.~Ionicioiu \cite{radu} have spotted a  couple of errors; we are most grateful to them all. We thank Ya. Belopolskaya,  L.~Carter, J.~Casti, G.~Chaitin, M.~Dinneen,
M.~Dumitrescu,  T.~Kieu, I.~Ibragimov, R.~Ionicioiu, A.~Lodkin, G.~P\u aun, J.~Summhammera, K.~Svozil, A.~Yafyasov   and
the anonymous referees for
heated and most inspiring discussions and criticism. Of course, nobody except the authors, are responsible for possible remaining errors.

\end{document}